\documentclass[aip,
 amsmath,amssymb,
reprint,%
]{revtex4-1}

\usepackage{subcaption}
\usepackage{graphicx}% Include figure files
\usepackage{dcolumn}% Align table columns on decimal point
\usepackage{bm}% bold math
\usepackage{orcidlink}
\usepackage{graphicx}
\usepackage{physics}
\usepackage{comment}
\usepackage[none]{hyphenat}
\usepackage{dcolumn}
\usepackage{bm}
 
\usepackage{blkarray}
\usepackage{amsmath,calc}
\usepackage{tikz}
\usetikzlibrary{mindmap}

\usepackage[utf8]{inputenc}
\usepackage[T1]{fontenc}
\usepackage{mathptmx}
\usepackage{etoolbox}

\makeatletter
\def\@email#1#2{%
 \endgroup
 \patchcmd{\titleblock@produce}
  {\frontmatter@RRAPformat}
  {\frontmatter@RRAPformat{\produce@RRAP{*#1\href{mailto:#2}{#2}}}\frontmatter@RRAPformat}
  {}{}
}%
\makeatother
\begin{document}

\preprint{AIP/123-QED}

%TC:ignore

\title[Finite-Time Braiding Dynamics within Topological Nanowire Qubits]{Finite-Time Braiding Dynamics within Topological Nanowire Qubits}
\author{Adrian D. Scheppe\orcidlink{0000-0002-5566-2744}}
\author{Michael V. Pak\orcidlink{0000-0002-3820-2789}}
 \email{michael.pak.4@af.au.edu}
\affiliation{Air Force Institute of Technology, 2950 Hobson Way, Wright-Patterson AFB, OH 45433}

\date{\today}

\begin{abstract}
Topological Quantum Computing has largely evolved towards a paradigm of manipulating edge localized Majorana within $p$-wave topological superconducting nanowires. To bridge the gap between physical qubit systems and quantum algorithms, we perform a dynamical analysis to extend what is known in the adiabatic regime, providing time-dependent gate elements for further qubit and algorithm modeling efforts. Our analysis covers dynamical considerations for two methods of shuttling domain edge bound Majoranas in a single nanowire system which both function by applying spatiotemporally dependent onsite and hopping parameters within the system's Hamiltonian. We then complicate this model by converting it into the T-qubit to calculate the finite-time gate representation of the shuttling techniques used in a more practical setting. These contributions provide insight for realistic experimental setups in the next-generation of qubit implementation and will hopefully facilitate fault tolerant scalable systems and universal gate design. 
\end{abstract}

\maketitle
%TC:endignore

\begin{quotation}
When working with quantum systems, one is consistently inconvenienced by the fact that they will always efficiently seek available pathways in which to decohere. We have contended with this abstract fact repeatedly as manifested in different forms, whether the question of the measurement mechanism, the emergence of classical reality, or the arrow of entropy. However, with the ever growing rise of new quantum technologies, we are being confronted by the most practical version of this question: Is it even possible to deny the artificial atom its natural inclination? Given nature's overwhelming persistance, it would be reasonable to give an emphatic negative response, but perhaps human persistance is on equal footing. Brutal pragmatism compels us to seek a ``good enough'' strategy to make the technology work in our favor. This is expected, but it should be no surprise. The list of such strategies is fairly short, and the challenges are legion.
\end{quotation}

As is well documented at this stage, the existence of topological quantum matter offers the only known path towards \textit{truely} fault tolerant observables which are protected from local noise by construction \cite{nayak2008non,freedman2003topological,stanescu2024introduction}. For Topological Quantum Computing (TQC) specifically, the general method bases logical qubits on non-local features \cite{alicea2011non, kitaev2001unpaired, chiu2015majorana}. For a well-known example, nontrivial band structure topology guarantees the emergence of boundary localized, zero energy states with an energy gap suppressing excitations into the bulk\cite{krantz2019quantum}. Of course, spatially bound states are cheap, occuring for crystal defects as well as superconducting (SC) phase discontinuities, so the elevated status of topological states comes from the fact that the physical organization of the system forces them to persist in the face of local environmental perturbations \cite{kitaev2009periodic,hasan2011three}. For some systems, these localized states carry non-abelian s tatistics with exchanges represented as noncommuting gates on the computational manifold \cite{ivanov2001non,aasen2016milestones}, and that possibility leads to the hope of fault tolerant quantum computing.

This is a promising scheme; however, in practice, the TQC recipe has its fair share of challenges \cite{aasen2016milestones}. Broadly, these issues are related to 1) state initiation, 2) gate operation, and 3) measurement protocol, manifesting in the forms of detection of Majorana Bound States (MBS) and verification of non-abelian anyonic statistics  \cite{yavilberg2019differentiating,zhang2018distinguishing,sau2011controlling}, non universality of braids \cite{bravyi2005universal, scheppe2022complete}, and fusion rule detection \cite{bartolucci2023fusion,mawson2019braiding,trebst2008short} respectively. Naturally, these challenges encourage researchers to relax certain requirements to make progress, but it is crucial to remind ourselves what is usually being sacrificed is the topological protection we sought out in the first place. Regardless, numerous TQC qubit platforms have been proposed, splitting into three paradigms: vortices \cite{ivanov2001non,posske2020vortex, scheppe2022complete,georgiev2006topologically}, chiral channels \cite{nilsson2008splitting,akhmerov2009electrically,aasen2020electrical,liu2022dynamical,lian2018topological}, and nanowires\cite{meidan2019heat,sau2011controlling,scheppe2025tight, alicea2011non, chiu2015majorana}, with nanowires taking center stage as the most widely investigated recently. This designation is due most likely to the fact that they are easiest to  model, fabricate, initiate, control, and measure compared to the former two options \cite{last2023key,sharma2022comprehensive,scheppe2023perturbing}. Relevant to this work, it is especially common currently to see $Y$ or $T$ configurations of nanowires\cite{alicea2011non,amorim2015majorana,tutschku2019topological}. 

One piece that has been missing from the literature on the subject is one that links the physical implementation of these T-qubits to the finite time, algorithmic implementation of gates within quantum codes. Specifically, how does MBS manipulation affect the finite-time dynamics of the ground state manifold? How do these actions correspond to computational gates? Answers to these questions are readily available for the infinite time regime; however, we supplement these works by building up a realistic model for the T-qubit and performing dynamical analysis to quantify time-dependent anyonic braiding. In Section \ref{sec:nanoDyn}, we cover the important details of Kitaev's chain by using numerical techniques on more rudimentary single and double nanowire systems. Then, we demonstrate the finite-time dynamics of MBS shuttling via potential and phase domain wall manipulation in Section \ref{sec:shuttle}. Lastly, in Section \ref{sec:tQubit}, we combine multiple nanowires together to calculate dynamical quantities for the $T$-qubit system using methods developed in Section \ref{sec:nanoDyn}. 

\section{Nanowire Dynamics: \label{sec:nanoDyn}}
\vspace{-10pt}
Kitaev's chain model is fundamental to any discussion of nanowire TQC platforms because it is predicted to host edge MBS \cite{chiu2015majorana,amorim2015majorana,lutchyn2010majorana,tutschku2020majorana}, so we provide a short description which covers the important details for our purposes. In practice, they are constructed by forcing semiconducting nanowires into an \textit{effectively} spinless $p$-wave SC phase by combining a strong spin-orbit coupling, an external parallel magnetic field, and an induced $s$-wave pairing potential from an adjacent SC material   \cite{alicea2011non,lutchyn2010majorana,oreg2010helical}. One can model a more realistic system which includes all terms mentioned, but we choose to model an $N$-site topological SC (TSC) nanowire with the following simplified BdG Hamiltonian,
\begin{equation}
    H= 
    -\sum^{N}_{n} \{\mu_n(t)\tau^{(n)}_z\hat{c}^\dagger_n\hat{c}_n+(\xi \tau^{(n)}_z+i\hat{\Delta}_n(t) \tau^{(n)}_y)\hat{c}^\dagger_{n+1}\hat{c}_n\}+h.c.,\label{eq:kitChain}
\end{equation}
where operators $\hat{c}^\dagger_n$ create a spinless fermion at site $n$ with onsite $\mu_n(t)$, $\xi$ is hopping, and $\hat{\Delta}_n(t)=\Delta_0 \text{exp}(\phi_n(t)\tau^{(n)}_z)$ is the pairing potential. The $2\cross 2$ matrices $\tau^{(n)}_i$ are the Nambu matrices acting on the particle \& hole degrees of freedom at site $n$. The ratio of parameters such that $|\mu/2\xi| < 1$ places the system into a topologically nontrivial phase. The dynamics of interest will be individually set by $\mu_n(t)$ and $\phi_n(t)$ for each of the coming systems, so, for now, we only set $\xi = 1$ eV and $\Delta_0 = .1$ eV for the remainder of this work.

One can analytically demonstrate the emergence of zero energy MBS by making the following parameter choices, $\Delta_0 =\xi$ and $\mu_n(t) = \phi_n(t)= 0$, then defining an initial onsite basis with $\hat{c}_n = \frac{1}{2}(\hat{\gamma}_{n}^A+i\hat{\gamma}_{n}^B)$\cite{kitaev2001unpaired,alicea2011non}. As a result, the Hamiltonian becomes, 
\begin{equation}
    H=-\frac{i\xi}{2}\sum_{n=1}^{N-1}(\tau_z^{(n)}+i\tau_y^{(n)})[\hat{\gamma}_{n+1}^A\hat{\gamma}_{n}^B-\hat{\gamma}_{n+1}^B\hat{\gamma}_{n}^A].
\end{equation}
Now define a new fermion basis by recombining $\gamma$ operators from adjacent sites such that $\hat{\gamma}^A_{n+1} =\hat{d}_n+\hat{d}_n^\dagger$ and $\hat{\gamma}_{n}^B = -i(\hat{d}_n-\hat{d}_n^\dagger)$ to transform $H$,
\begin{align*}
\label{eq:mbsKC}
    H=-\frac{\xi}{2}\sum_{n=1}^{N-1}(\tau_z^{(n)}+i\tau_y^{(n)})&[(\hat{d}_n+\hat{d}_n^\dagger)(\hat{d}_n-\hat{d}_n^\dagger)\\&-(\hat{d}_{n+1}-\hat{d}_{n+1}^\dagger)(\hat{d}_{n-1}+\hat{d}_{n-1}^\dagger)].
\end{align*}
Neglecting next nearest neighbor interactions, we are left with the following,
\begin{equation}
    H=-\xi\sum_{n=1}^{N-1}(\tau_z^{(n)}+i\tau_y^{(n)})[\hat{d}_n^\dagger\hat{d}_n-\frac{1}{2}]\label{eq:mbsKC},
\end{equation}
where we use fermionic brackets $\{d_n,d^\dagger_{n'}\} = \delta_{nn'}$. Since the sum begins at $n=1$ and ends at $n=N-1$, operators $\hat{\gamma}^A_1$ and $\hat{\gamma}^B_N$ are left out of the summation, implying two zero energy states $\ket{0}:=\ket{\psi_0}$ and $\ket{1}:=d_0^\dagger\ket{0}$ such that $\hat{d}_0 = \hat{\gamma}^A_1+i\hat{\gamma}^B_N$ and $d_0\ket{1}=\ket{0}$. Based on the form of Equation \ref{eq:mbsKC}, one can infer that non-zero couplings between these edge modes could be modelled by a perturbing Hamiltonian of the form\cite{posske2020vortex},
\begin{equation}
\label{eq:mbsPert}
    H_p = i\Gamma(t)\hspace{1pt} \hat{\gamma}^A_{0}\hspace{1pt}\hat{\gamma}^B_{N}.
\end{equation}  
For brevity, we set $\hat{\gamma_1}:= \hat{\gamma}^A_0$ and $\hat{\gamma_2}:= \hat{\gamma}^B_N$. In what follows then, our computational system is modelled with gate interaction $H_p$ and basis defined as $\{\ket{0}, \ket{1}:=\hat{d}_0^\dagger\ket{0} \}$ for intial states that are less/ greater than zero upon fusion of the MBS. 

\vspace{-10pt}
\subsection{Single Nanowire\label{sec:single}}
\vspace{-10pt}
A single nanowire system is too simple for useful 1D particle exchanges and is thus not practical. However, we present this simplest case to demonstrate how the analytic and numeric calculations relate to one another as well as to demonstrate the core aspects of our finite-time calculation. Toward that goal, one can couple two edge MBS via an appropriate perturbing Hamiltonian. For example, by manipulating voltage \textit{globally} such that particle-hole symmetry is broken with respect to time, one can lift the ground state degeneracy and develop a nonzero Equation \ref{eq:mbsPert} with $\Gamma(t) = \eta t$. This in turn generates dynamics for our computational basis via, $U(t_f,t_i) = \text{exp}(1/i\hbar \int^{t_f}_{t_i} dt' H_p(t'))$. To see this, use the computational representation for the MBS operators,
\begin{gather}
\gamma_1 =\tau_x =\begin{pmatrix}
    0&1\\
    1&0
\end{pmatrix},\hspace{10pt}\gamma_2 =\tau_y= \begin{pmatrix}
    0&-i\\
    i&0
\end{pmatrix},
\end{gather}
such that $H_p = i\eta t\tau_x\tau_y = -\eta t\tau_z$,
\begin{gather}
\label{eq:naiveGate1}
    U = \begin{pmatrix}
        e^{-i\phi(t)}&0\\
        0&e^{i\phi(t)}
    \end{pmatrix},
\end{gather}
for $\phi(t) = \frac{\eta}{2\hbar} t^2$. This solution implies that the perturbation simply rotates the basis' phases indefinitely, but note that a symmetry breaking $H_p$ increases the gap between ground states linearly with time. Eventually, the  in-gap states will meet the bulk spectrum, and, at that point, bulk states cannot be neglected within our analytical model\cite{bravyi2011schrieffer}.
\begin{figure}[t]
    \centering
    \includegraphics[width=\columnwidth]{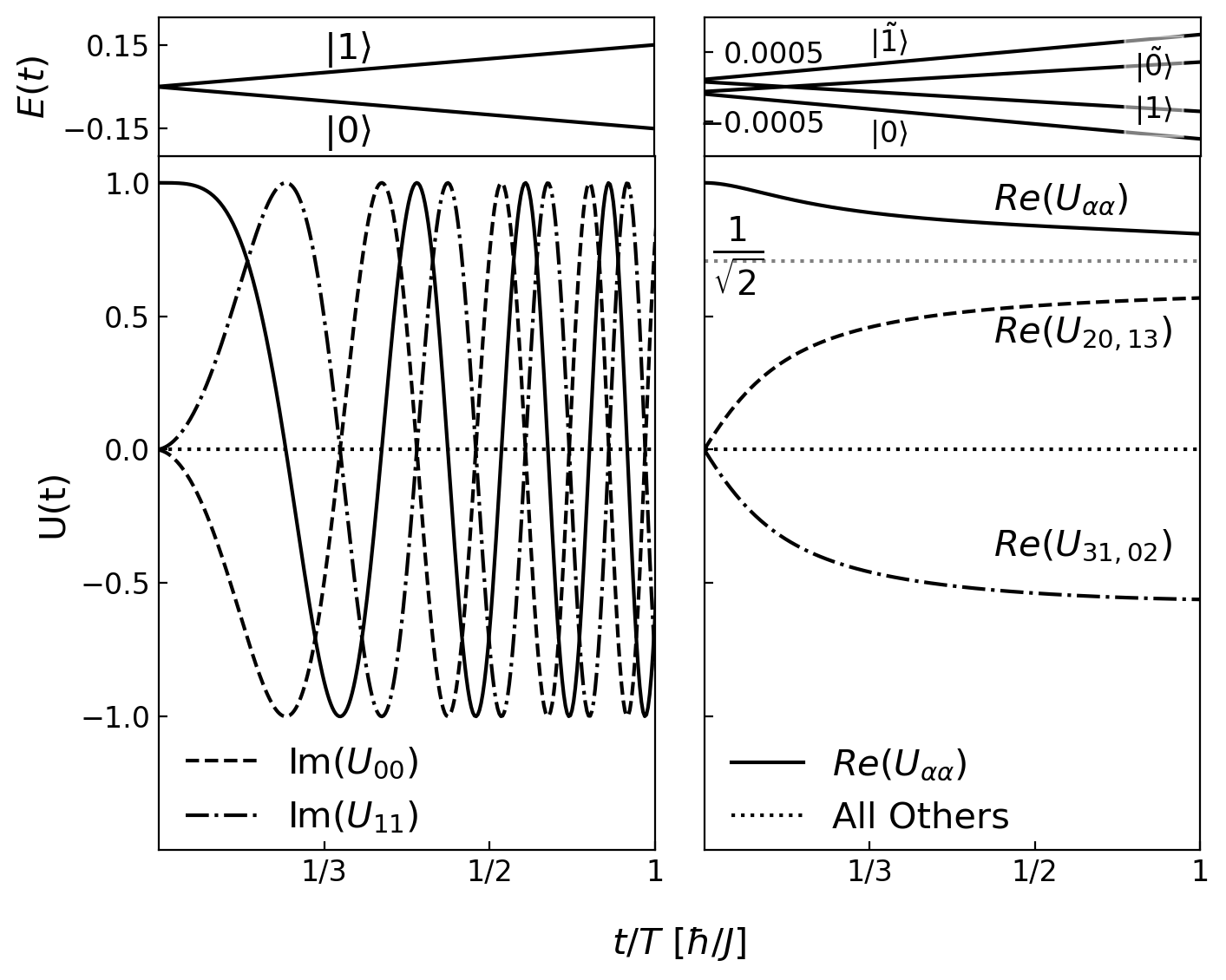}
    \caption{Simple System Finite-Time Gates. Left plot depicts $U(t)$ produced by the P-H symmetry breaking perturbation for a single nanowire. The right plot depicts similar information for the time-dependent coupling of two nanowires. For both scenarios, the plots of $E(t)$ imply that these perturbations change the content of the ground state manifold in such a way to spoil fault tolerance.}
    \label{fig:naiveGates}
\end{figure}

As a validation exercise, we reproduce this perturbation within Equation \ref{eq:kitChain}. For $H_{tot}(t) = H+H_p(t)$, a sufficient particle-hole symmtery breaking term can be written as,
\begin{equation}
    H_p(t) =\eta t \sum_n^N \tau^{(n)}_x\hat{c}^\dagger_n\hat{c}_n,
\end{equation}
with $\eta = 5* 10^{-4}$. First, we treat time as a parameter to acquire a dimension $2N$ basis by solving $H_{tot}(t_i) \psi^\alpha_i = \varepsilon^\alpha_i \psi^\alpha_i$ for each $t_i$. We use subscripts to emphasize that these are not the dynamical quantities yet, but rather a convenient moving basis to quantify gate operations on the ground state manifold. These can equally well be named the adiabatic solutions, because, for slow enough potentials, they do in fact resemble true dynamics of the system. A moving basis is unecessary at this stage, but we will need it for Sections \ref{sec:shuttle} and \ref{sec:tQubit}. So, we preemtively build up the method here for use later as well. 

To numerically propagate the system from time $t_i$ to $t_{i+1}$, one uses, 
\begin{equation}
\label{eq:adiabatic}
\ket{\psi(t_{i+1})} =  e^{\frac{1}{i\hbar}H(t_i)\delta t}\ket{\psi(t_i)},
\end{equation}
with parenthesis emphasizing dynamics now. We denote full time propagation as $\ket{\psi(t_i)} = U(t_i,0)\ket{\psi_{\beta,0}}$ for some initial ground state $\ket{\psi_{\beta,0}}$. The time dependent gate elements, 
\begin{equation}
    U_{\alpha\beta}(t_i) = \bra{\psi_i^{\alpha}}U(t_i,0)|\psi^\beta_{0}\rangle\nonumber,
\end{equation}
of $U(t)$ govern transitions from the initial $\beta^\text{th}$ ground state into the $\alpha^\text{th}$ ground state at time $t_i$. For a single $N = 50$ nanowire's 2D ground state manifold, we set $\delta t = 1$ and total time, $T=300$, to calculate $U_{\alpha\beta}(t)$, plotting results on the left side of Figure \ref{fig:naiveGates}. The time dependent behavior is nearly exactly reproduced by a sinuisodal trajectory with a $\phi\propto t^2$ which is opposite in sign based on initial state and with off-diagonal elements equal to zero. Note also the increasing energy gap between $\ket{0}$ and $\ket{1}$ plotted as well, echoing our previous statements that this gate will eventually involve bulk elements as well.

\vspace{-10pt}
\subsection{Double Nanowire\label{sec:double}}
\vspace{-10pt}
For two uncoupled TSC, the 4D ground state can likewise be specified in terms of four delocalized fermions or eight edge bound $\gamma_j$'s. Analytical behavior in this case is modelled by a tunneling Hamiltonian, 
\begin{equation}
\label{eq:tunnelHam}
    H_T = \Gamma(t) (d^\dagger_{L}d_{R}+d^\dagger_{R}d_{L}),
\end{equation}
which couples the left and right systems. Define $\gamma_{2,3} := d_{L,R}+d_{L,R}^\dagger$ to be MBS at the center of the total system and $\gamma_{1,4} := i(d_{L,R}-d_{L,R}^\dagger)$ on the outer edges. After analogous math to the single nanowire case, Equation \ref{eq:tunnelHam} becomes,
\begin{equation}
    H_T = -\frac{i}{2}\Gamma(t) \gamma_2 \gamma_3.
\end{equation}
Proceeding carefully, one defines the computational states as $\{\ket{00},d_1^\dagger \ket{00}, d^\dagger_2 \ket{00}, d_1^\dagger d_2^\dagger \ket{00}\}$, such that $\gamma_2\gamma_3 = i \tau_y \otimes\tau_x$ which follows from fermionic anticommutation brackets.

Let $\Gamma(t) = \epsilon t$ again so that time propagation is,
\begin{equation}
    U(t) = e^{i\phi(t) \tau_y\otimes\tau_x },
\end{equation}
resulting in the following gate,
\begin{equation}
    U(t) = \text{cosh}\phi(t)\mathbb{I}\otimes \mathbb{I} +i\text{sinh}\phi(t)\tau_y\otimes\tau_x
\end{equation}
In the limit as $t\rightarrow\infty$, this approaches the expected adiabatic value of $U(\infty) =\frac{1}{\sqrt{2}} (\mathbb{I}\otimes\mathbb{I}
+i\tau_y\otimes\tau_x)$\cite{scheppe2022complete}. 

Concluding the demonstration, we present the same calculation to the single nanowire case using identical system parameters as well as the same values of $\delta t$ and $T$. The comparable numerical model includes a time dependent $H_p$ much like the second term within Equation \ref{eq:kitChain} which connects the two side-by-side systems over time with coupling $\eta = 10^{-5}$. We plot the resulting $U(t)$ on the righthand side of Figure \ref{fig:naiveGates}. Notice first that the absolute value of each element do in fact approach the analytic value of $1/\sqrt{2}$. We can also see that there are two positive and negative off-diagonal elements which is consistent with the analytic gate as well. Referencing the right energy subplot now, the totally uncoupled system at $t=0$ has two degenerate levels which correspond to two 2D manifolds spanned by $\ket{0}, \ket{\tilde{0}}$ and $\ket{1},\ket{\tilde{1}}$ native to each nanowire. Coupling causes an energy split between these levels, but notice the elements of $U(t)$ only states which come from the correct parity sector connect. 
\begin{figure}[t]
    \centering
    \includegraphics[width=\columnwidth]{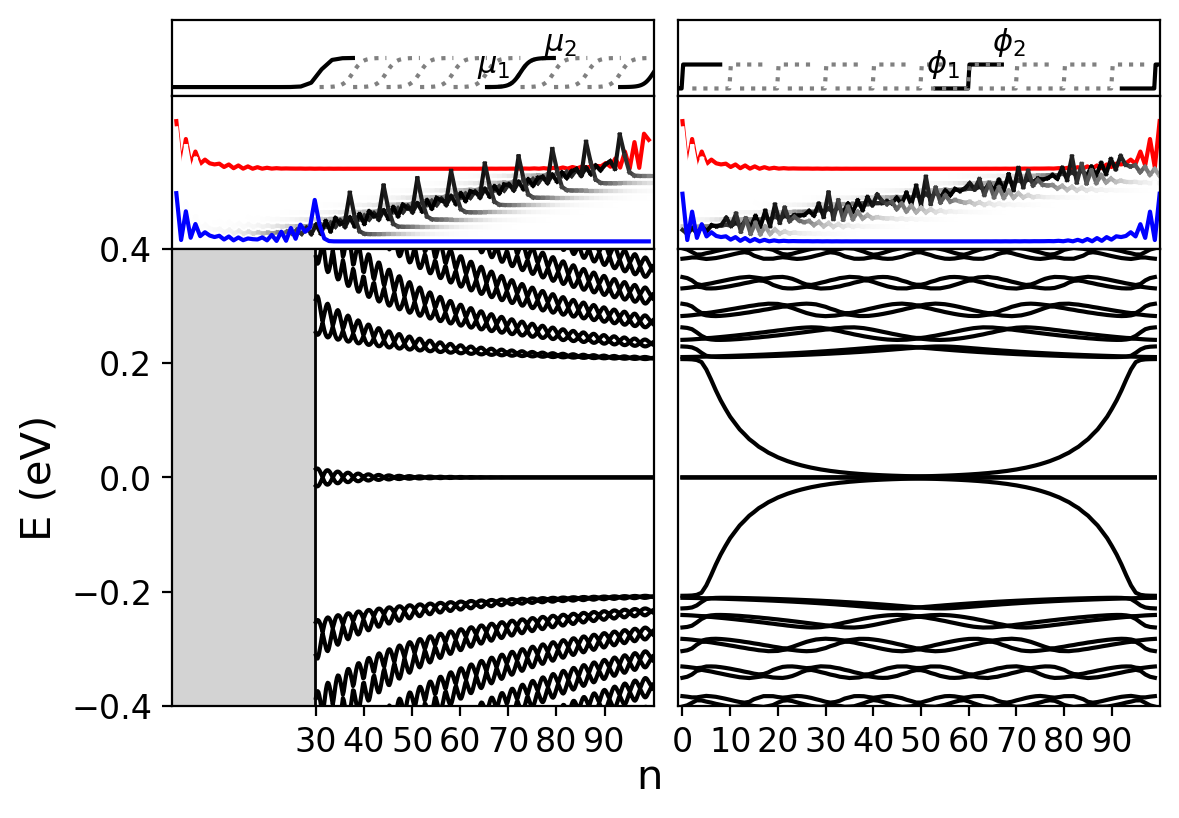}
    \caption{Adiabatic Shuttling. Left plot depicts the spectrum for adiabatic $\mu-$shuttling by shifting a domain wall sigmoid potential profile along the chain. Right plot depicts identical calculation for $\phi-$shuttling for a travelling phase discontinuity. We use the associated eigenstates as a moving basis in our time-dependent calculation.}
    \label{fig:adiabaticShuttle}
\end{figure}

\vspace{-10pt}
\section{Shuttling MBS\label{sec:shuttle}}
\vspace{-10pt}
The previous gates end up directly affecting the structure of the ground state manifold by either changing the number of basis elements or entirely destroying its existence over time. So, while simpler for demonstrative purposes, they are limited for any practical computations. 

Two better techniques have appeared in the literature which both function by manuevering the domain boundary of the topological system by means of spatiotemporal variation of system parameters. Alicea \textit{et al}\cite{alicea2011non} outlined the first method where $\mu_n(t)$ in Equation \ref{eq:kitChain} is set to be within the topological range of values for a section of the nanowire while the rest is trivial. One can imagine using voltage nodes to specify $\mu_n(t)$ on certain lattice sites to switch on/off topological nature in specific regions, thereby shuttling the edge MBS along the wire. Similarly, the second technique outlined by Chiu \textit{et al} \cite{chiu2015majorana} demonstrates that a moving discontinuity in $\phi_n(t)$ is likewise a valid method for shuttling bound states. Instead of voltage, the physical implementation is thought to be an onsite magnetization which carry a bound state completely across the system. Once completed the total action recreates a braid protocol on the ground state manifold, providing an option for 1D braiding. Henceforth, for convenience, we will name the first and second method the $\mu-$ and $\phi-$ methods respectively. Our core contribution to this lineage of work is a dynamical calculation for these two shuttling methods. 

We achieve this by modeling both quantities as a sigmoid profile over sites $n$,
\begin{equation}
\label{eq:sigmoid}
    f_n(t) = -f_1 + \frac{f_1 - f_2}{1+e^{-\sigma(n-t)}},
\end{equation}
which has a step centered at $t$ and where $f_{1,2}$ are values of the parameters on the left/right of the sigmoid step and smoothness determined by $\sigma$. With that, we essentially perform the same calculations from Section \ref{sec:nanoDyn}, finding first the moving basis for Equation \ref{eq:kitChain} while incorporating Equation \ref{eq:sigmoid}. The results of the adiabatic shuttling calculation are depicted in Figure \ref{fig:adiabaticShuttle}. These are made for an $N = 100$ system by setting $\mu_1 = .185$ eV and $\mu_2 = 4$ eV for the $\mu$-method. Likewise, we set $\phi_1 = 0$ and $\phi_2=\pi$ for the $\phi-$method. We let $\sigma = 1$ to set the smoothness of the $\mu$-sigmoid; however, the $\phi-$sigmoid's resultant bound states are highly dependent on the sharpness of the changes in SC phase. This is due to the fact that the shuttling bound state is an Andreev bound state first, then MBS if well seperated enough from the boundaries of the system \cite{scheppe2025tight}. This explains why the spectrum becomes degenerate once the discontinuity reaches the mid point of the system in Figure \ref{fig:adiabaticShuttle}. Naturally, gap closure is not ideal for computations since the approaching bulk states cause scattering from the ground state. Therefore, we address a sharp $\phi$-sigmoid as well as a smoother $\sigma = 1$ sigmoid to explore this parameter space. Lastly, the calculations for this section are made by setting $T=1000$ and $\delta t = 1$.

\vspace{-10pt}
\subsection{$\mu-$Method\label{sec:muShuttle}}
\begin{figure}[t]
    \centering
    \includegraphics[width=\columnwidth]{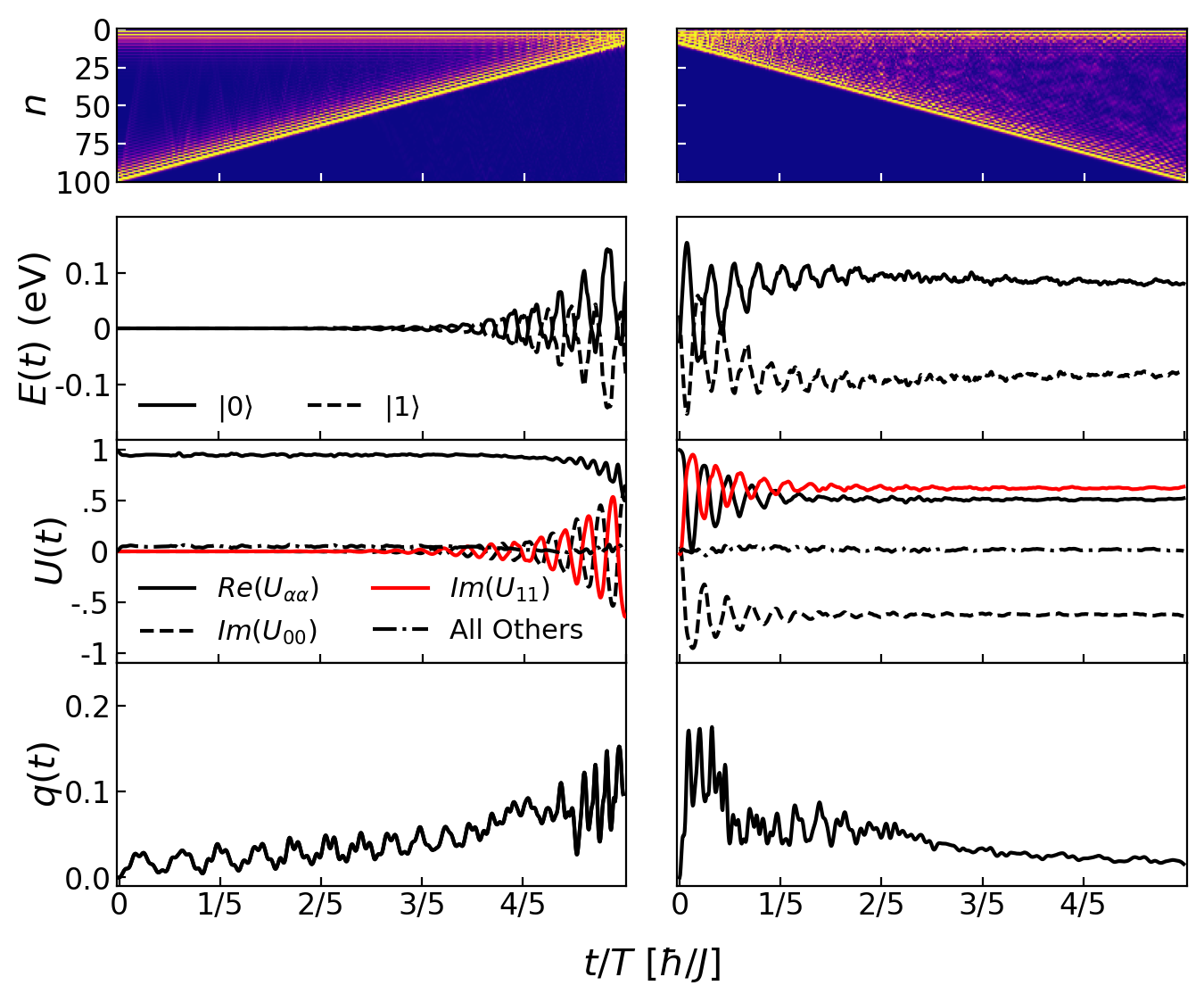}
    \caption{Finite-Time $\mu-$Shuttling. Left column depicts the time dependence of a MBS being $\mu-$shuttled towards one another, and the right column shows comparable information for MBS moving away from one another. To visualize this trajectory, we plot $\rho(t)$ in row one. We calculate $E(t)$ in row two to compare how the energy gap of the initial configuration affects energetics in future states. $U(t)$ is nearly identity for all time in the far-out scenario, while the close-in scenario leads to a complex rotation as well as significant scattering into the bulk as indicated by $q(t)$.}
    \label{fig:timeDepShuttles_mu}
\end{figure}

\vspace{-10pt}
There are two scenarios for the $\mu-$shuttling system that are of interest. Both are featured on the left and right columns of Figure \ref{fig:timeDepShuttles_mu} respectively. The left depicts initial conditions such that $|\psi(0)\rangle$ is localized to the $n=0,100$ sites, i.e. the two MBS start far apart and are brought into one another. On the right, the MBS initiate close-in at $n=0,30$ and are sent away from one another. Using Equation \ref{eq:adiabatic}, the states are propagated forward in time to produce a time-dependent local density,
\begin{equation}
    \rho(t) = \bra{\psi(t)}\bigotimes^N_n \mathbb{I}^{(n)} \ket{\psi(t)}.
\end{equation} 
We plot this in the top row of Figure \ref{fig:timeDepShuttles_mu} to visualize the state's movement in time. Qualitatively, both scenarios retain domain edge bounded MBS, but the initially close-in system shows notably more distortion in the density. This occurs because the initial configuration  significantly splits the degeneracy in the ground state more than the far-out scenario. We make this point more clear comparing $E(t) = \bra{\psi(t)} H \ket{\psi(t)}$ for each scenario on the second row. Labelling each curve with their initial state, one can see that the far-out configuration leads to a more stable low energy manifold while close-in MBS generate early oscillations in the states, spoiling the possibility of the system settling into a zero energy manifold in the future. The resulting elements of $U(t)$ are plotted next on the third row. From the magnitude of the diagonal elements $U_{\alpha\alpha}$ compared to off-diagonal, we can have high confidence that the far-out system will remain in its initial state until the MBS are approximately 25 sites away from one another. This is indicated by the oscillations in the complex part of the element at $t/T=4/5$. The close-in system's off-diagonal elements also remain zero, but $|U_{\alpha\alpha}|^2\cong .683$ as well, meaning the state has significiantly scattered into the bulk. To quantify this, we introduce,
\begin{equation}
    q(t) = \sum^{k}_{\gamma\notin g.s.m} |\langle\psi^\gamma_i|\psi(t)\rangle|^2,
\end{equation}
the probability of the dynamic state scattering into a state outside the groundstate manifold for $k$ closest bulk bands. This quantity is plotted in the final row of Figure \ref{fig:timeDepShuttles_mu} to show that the scattering is indeed significant at the start of the close-in system, indicating that much of the system will decohere in this sense right at the start. Alternatively, the far-out scenario performs well until the end where the MBS collide and fuse.

\vspace{-10pt}
\subsection{$\phi-$Method\label{sec:phiShuttle}}
\vspace{-10pt}
The main complication for shuttling states using $\phi_n(t)$ is that there exists approaching bulk states that appear for any realistic length of TSC wire. The system can be made to be gapless for all values of $M$ if made short enough\cite{chiu2015majorana}, but it would be most  useful at this stage to introduce strategies to contend with the longer system so as to make commentary about a more realistic setting. The sharper sigmoid profile is no doubt the intended setting for the 1D braiding protocol, but, by softening the phase difference, one ruins the strict domain seperation between left and right regions and subsequently lifts the degenerate bands. 

\begin{figure}[t]
    \centering
    \includegraphics[width=\columnwidth]{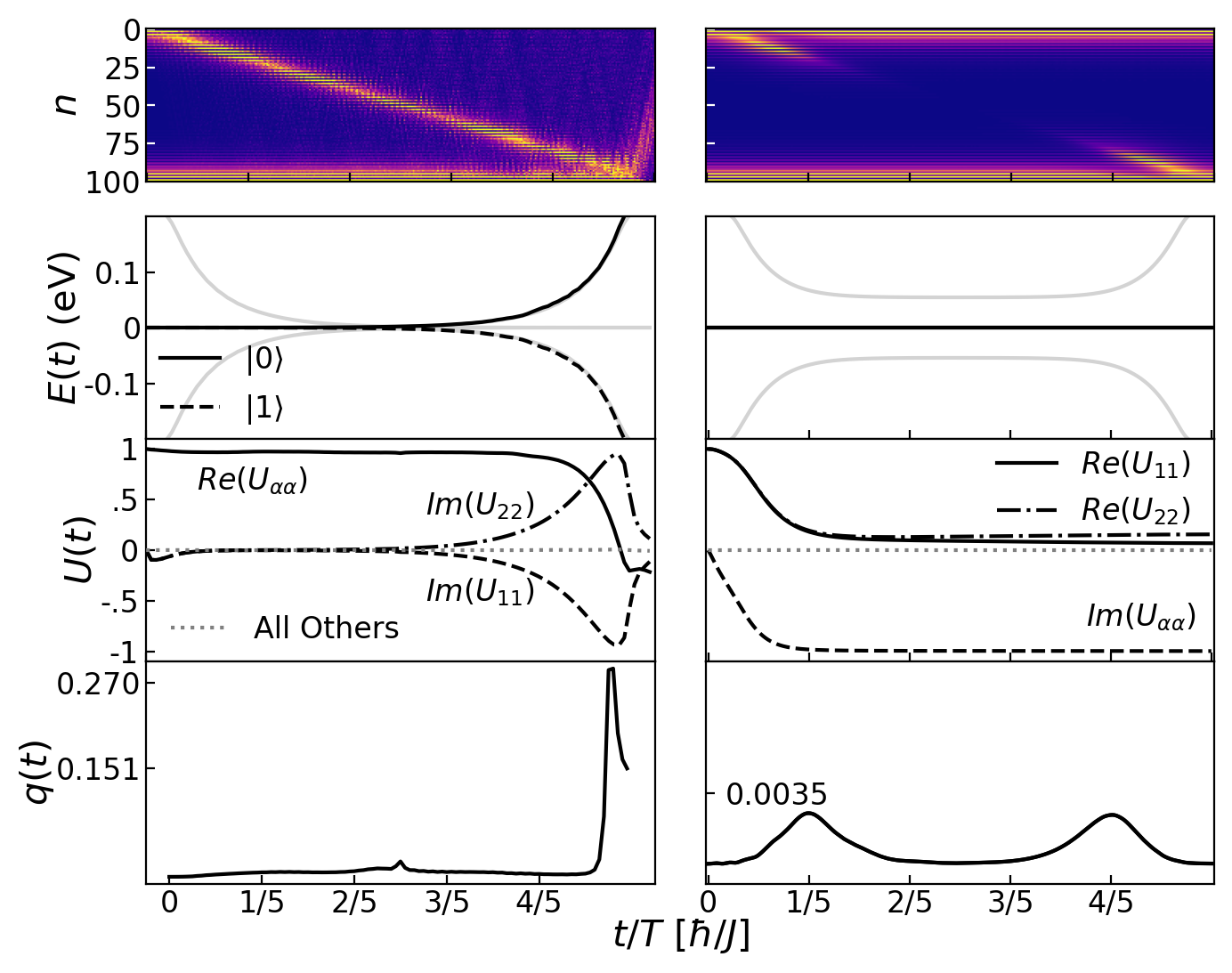}
    \caption{Finite-Time $\phi-$Shuttling. Left column depicts the relevant dynamic quantities for the 1D braid with a step-like profile and the right depicts a softened $\phi-$sigmoid system. Here we see the gapless nature of the sharp profile system spoils the protected status of the ground state manifold as indicated by the rising $E(t)$ and sharp increase in $q(t)$. The softened system maintains its ground state manifold for all values of $t$.}
    \label{fig:timeDepShuttles_phi}
\end{figure}

Following our recipe once again, $\rho(t)$ for both scenarios are plotted in the first row of Figure \ref{fig:timeDepShuttles_phi}. The sharper system on the left shows qualitatively clear bounded character for all time while the softened system seemingly disappears throughout the middle portions of the process. Regardless, within the second two, we plot $E(t)$ in black alongside the adiabatic values in light grey to compare the gap versus gapless scenarios. The gapless system has both initial states gaining/ losing energy as they pass through $t/T=1/2$ whereas the softened system has a gapped ground state manifold for all $t$. The third row indicates that $U(t)\cong \mathbb{I}$ for $t/T<1/2$, but the sudden increase in the complex part of these diagonal elements indicates a momentary $\pm\pi/2$-phase rotation of the state after roughly $t/T\cong 9/10$ before dropping to nearly zero, accompanied by a sudden increase in $q(t)$. Likewise, the softened scenario rotates by an equal amount entirely before $t/T=1/5$, but remains stable at that value for all $t$. The dramatic scattering does not occur for the softened potential where $q(t)$ remains quite small, implying that the time-dependent potential's effect is fault tolerant

\vspace{-10pt}
\section{T-Qubit: \label{sec:tQubit}}
\vspace{-10pt}
In this final section, we apply our calculations to a more interesting and practical system of nanowires mentioned in the introduction. The T-qubit system is fairly popular within the literature, so it is a natural first step to extend the model and methods of Section \ref{sec:nanoDyn} by attaching one more nanowire to the midpoint of Equation \ref{eq:kitChain}. For our calculation, we let the top portion and bottom portions of the system be 51 and 25 sites long respectively. The lower nanowire is attached to the $n=26$ sites of the upper one.

\begin{figure*}
\begin{subfigure}{.4\textwidth}
  \centering
  \includegraphics[width=\linewidth]{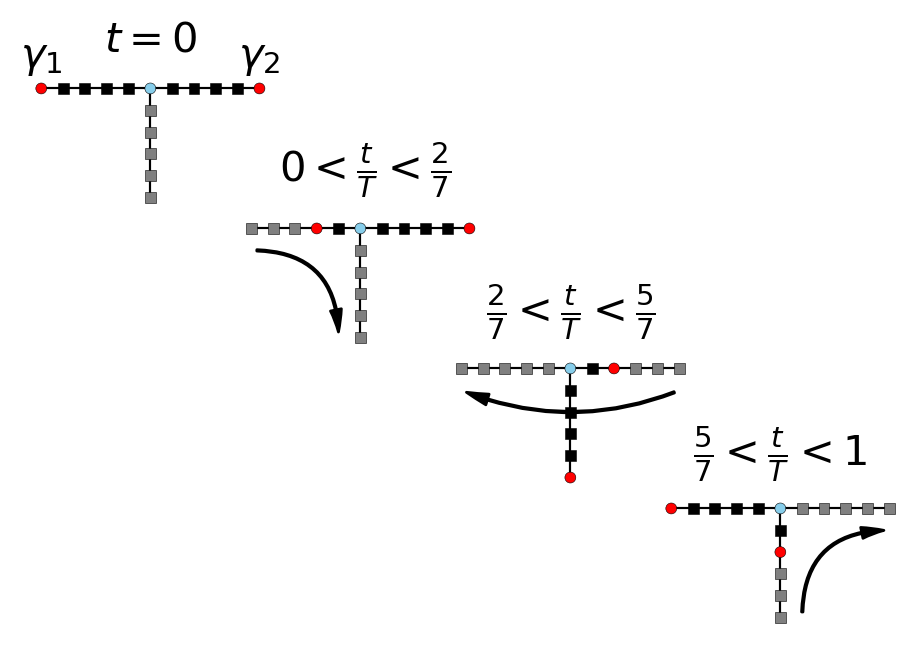}
  \caption{}
  \label{fig:muprotocolDiag}
\end{subfigure}%
\begin{subfigure}{.4\textwidth}
  \centering
  \includegraphics[width=\linewidth]{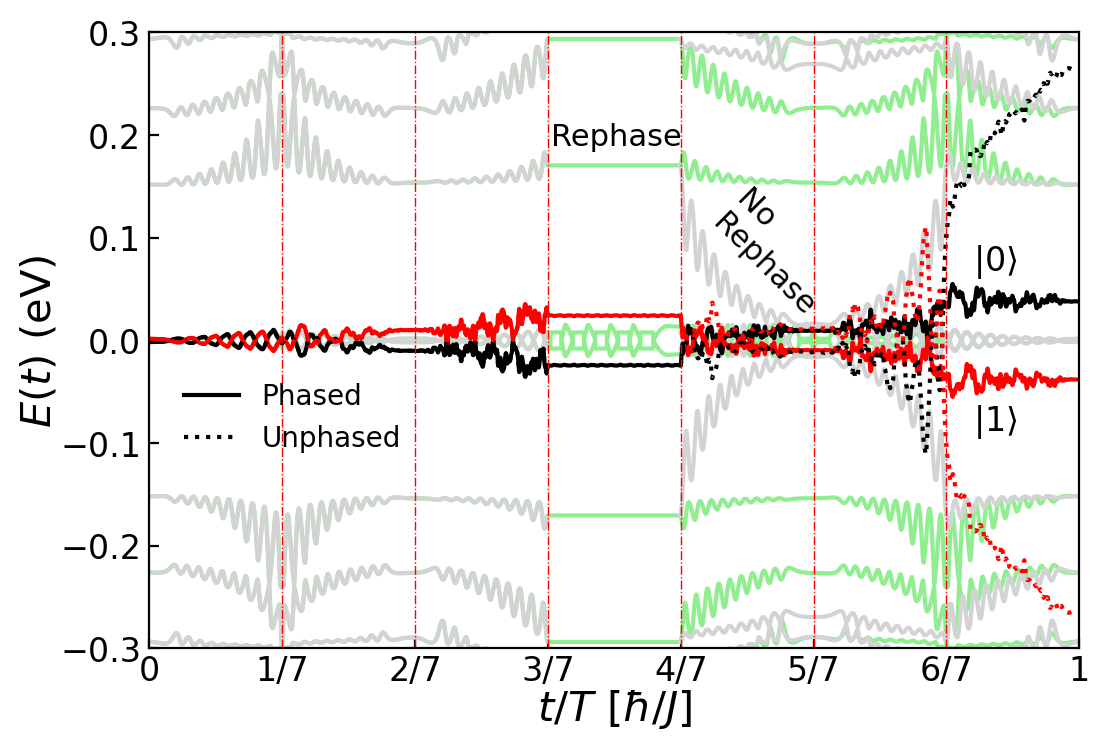}
  \caption{}
  \label{fig:muprotocolEnergy}
\end{subfigure}\\
\begin{subfigure}{.4\textwidth}
  \centering
  \includegraphics[width=\linewidth]{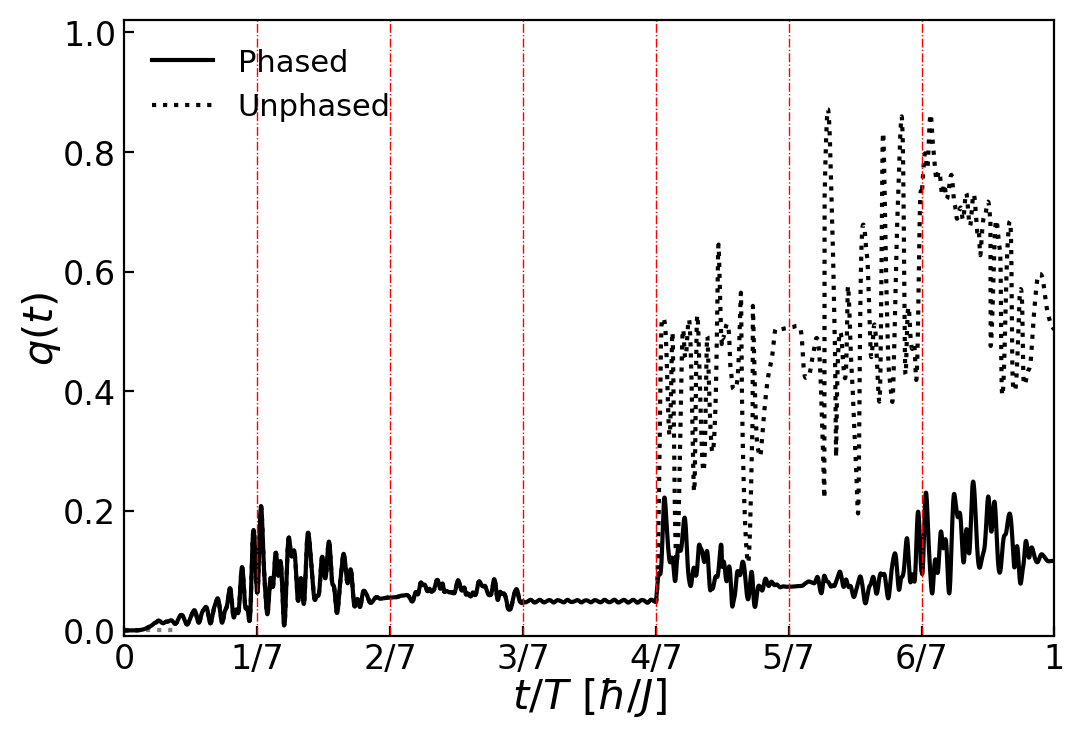}
  \caption{}
  \label{fig:muprotocolDeco}
\end{subfigure}
\begin{subfigure}{.4\textwidth}
  \centering
  \includegraphics[width=\linewidth]{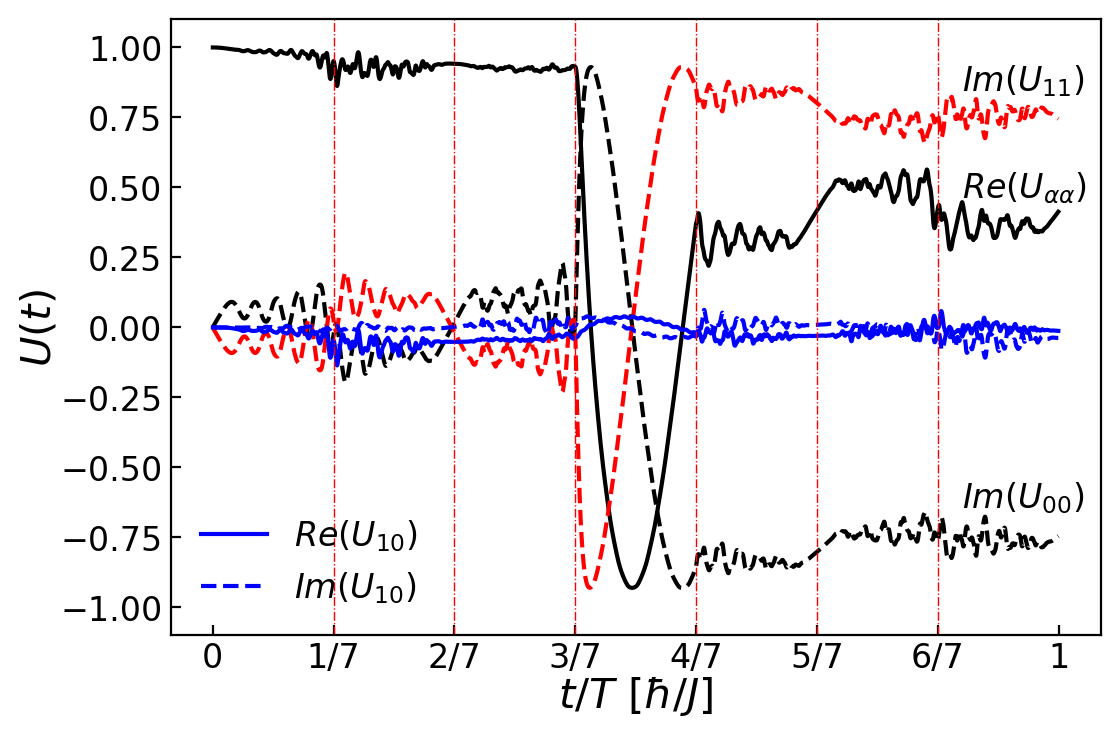}
  \caption{}
  \label{fig:muprotocolGate}
\end{subfigure}
\caption{$\mu-$protocol. (a) Diagram of the braiding action used in this calculation. The grey (black) sites are topologically (non)trivial, and MBS sit at the red sites. While the right MBS moves across the junction at $t/T = 3/7$, we rephase the lower branch of the system to keep the bulk states from approaching the ground state manifold. (b) We plot the adiabatic energy spectrum for the unphased and phased system in light grey and green respectively. Notice, the bulk state which enters the gap after $t/T=4/7$. We also plot the dynamic $E(t)$ in solid black and red. (c) Computing $q(t)$ demonstrates the effect that the junction bound state has on dynamics. (d) $U(t)$ is approximately diagonal with a nontrivial complex rotation on the computational manifold.}
\label{fig:tqbMu}
\end{figure*}

The specifics of the model are very similar to the dual nanowire system considered in Section \ref{sec:double} with the caveat that the creation/annihilation operators of Equation \ref{eq:tunnelHam} tunnel electrons from one end of a nanowire to a middle site of the other. This architectural choice converts the 1D platform into a 2D one with more shuttling paths. The motivation for this change is that MBS can potentially be $\mu-$shuttled around one another using the additional chain\cite{alicea2011non} as depicted in the top most graphic of Figure \ref{fig:tqbMu}. There, the grey (black) sites indicate the region of topological (non)triviality, and the red sites indicate localized MBS. To model the now 2D time-dependent potential, we multiply four sigmoids similar to Equation \ref{eq:sigmoid} which contain independent parameters to specify a rectangular region of topological nontriviality. Over time, this function scans aross the system to execute the braiding protocol depicted in Figure \ref{fig:muprotocolDiag}. Aside from these new additions, all other parameters are identical to the previous models. 

\vspace{-10pt}
\subsection{$\mu-$protocol}
\vspace{-10pt}
Prior to showing our results for this protocol, we must discuss one important technical aside. It is known that attaching two nontrivial TSCs in this manner introduces  extra low lying states into the gap\cite{tutschku2020majorana,stanescu2018building,huang2018quasiparticle,scheppe2025tight}. Namely, two bound states appear which are localized to the junction point, depicted as the green circular site in Figure \ref{fig:muprotocolDiag}. However, it is also the case that even for an "L" shaped system one can potentially have these bound states appearing. This occurs because one is implicitly choosing a directionality when populating the tight-binding model. To be clear, once one has specified the upper system from left to right using Equation \ref{eq:kitChain}, the lower system can be created starting at the bottom or at the top. Whichever direction one chooses effectively places the bottom portion of the system $\pi$ out of phase with either the left or right side of the upper system. Thus, if one is not careful, even an "L" shaped system can potentially have a phase discontinuity and bound state localized at the bend. 

This is important for our considerations, because, during the $\mu-$protocol, the system will temporarily pass through a right/left facing "L" configuration at $t/T = 2/7$, $5/7$ respectively. This means we will need to contend with the bound states that appear. To maintain the gap for the entire braid, we find that rephasing the middle nanowire at key moments of time can reorient the implicit directionality of the lower system and maintain the gap. After setting $\delta t = 1$ and $T = 3500$, we report both the adiabatic dispersion and fully dynamic $E(t)$ in Figure \ref{fig:muprotocolEnergy}. The light colored plots are the adiabatic values where we compare the unphased braid (grey) to the phased (green). To compare on equal footing, we have also shifted the unphased system to align the dispersion curves properly. Without a rephasing step at $t/T = 3/7$, the junction bound bulk state enters the gap. No such state appears for the rephased braid, and we find a stark difference for $E(t)$ for the computational manifold as indicated by the difference between dotted and solid curves. Calculation of $q(t)$ in Figure \ref{fig:muprotocolDeco} quantifies the effect that a rephase step has on this scattering process, showing that the phased system tempers bulk scattering by keeping it roughly below .2 for all time. By $t/T = 4/7$, the unphased system will have surely scattered into the bulk, ruining computational prospects. 

Lastly, we showcase the $2\cross 2$ $U(t)$ elements for the phased system  in Figure \ref{fig:muprotocolGate}. Similar to the previous braids, this protocol has its off-diagonal elements close to zero for all $t$. By the $t/T=1$ point, this braid differentiates itself from previous systems by producing $U(t)$ such that the ground states are phased differently from one another. Before the bound state reaches the mid point at $t/T<3/7$, the system responds similarly to the single nanowire case in Section \ref{sec:muShuttle}, i.e. $U(t)\cong\mathbb{I}$. The rephasing step at $t/T = 3/7$ generating oscillations in the elements of $U(t)$, constituting the the primary effect of the braid. For a single topological qubit, these results are consistent with the predicted gate operation which follows from adiabatic manipulation of two MBS\cite{scheppe2022complete}. That is, swapping anyons defined within the same fermion produces a phase gate on the qubit. For this particular choice of $T$, we have arrived at a pleasing result.

\vspace{-10pt}
\subsection{$\phi-$protocol}
\begin{figure}[t]
    \centering
\includegraphics[width=\columnwidth]{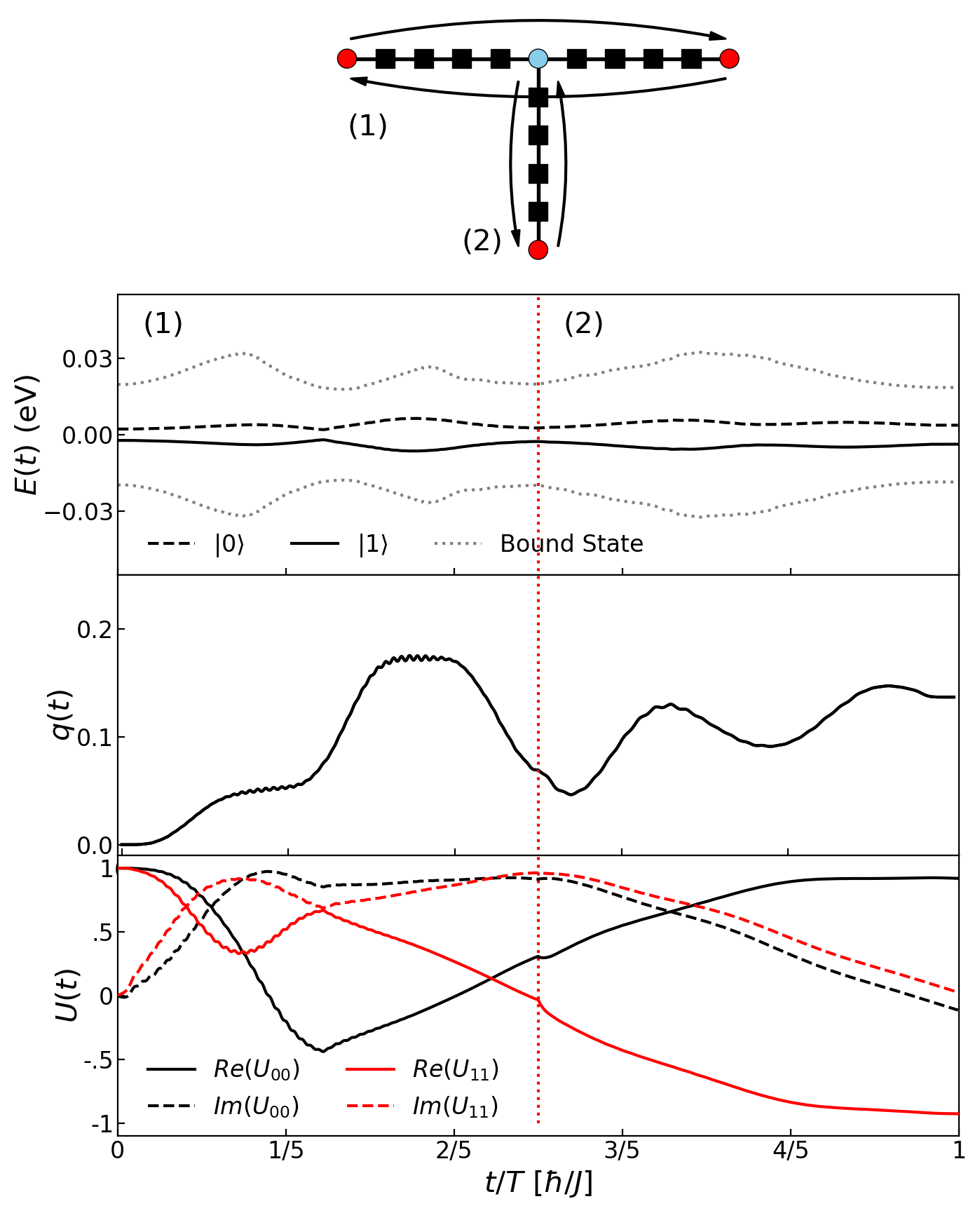}
    \caption{$\phi-$protocol. We scan a $\phi$ discontinuity from left to right along the top part of the fully topological system then from top to bottom along middle portion. The plot of $E(t)$ shows that there are four states within the bulk, two edge bound and two junction bound. Even still, the bands remain distinct and seperated throughout the braid, indicating that this braid is well-behaved. We include these extra states in our calculation of $q(t)$ to reinforce our point since scattering into non computational states remains less than .2 for all $t/T$. The total effect of $U(t)$ is effectively the phase gate.}  
    \label{fig:TQb_phiprotocol}
\end{figure}

\vspace{-10pt}
As a final calculation, we address how the $\phi-$method might be implemented within the T-qubit in a way which makes use of the extra branch. Placing the system into an initial configuration much like the first step of the $\mu-$protocol, one could reproduce identical results to what is reported in Section \ref{sec:phiShuttle}. This would be a redundant computation, so we wish to deviate from the last subsection slightly. Instead, we place the entire system into the topological regime with MBS on all three points of the qubit and implement the protocol depicted in Figure \ref{fig:TQb_phiprotocol}. With this initial configuration, we choose to scan a $\phi-$sigmoid left to right along the upper portion of the system in the first step, then top to bottom in a subsequent step. This is effectively using the 1D braid twice on the horizontal, then vertical parts of the system.

Setting $\delta t = 1$ and $T = 1000$, we report the results of this calculation in the remaining subplots of Figure \ref{fig:TQb_phiprotocol} with $E(t)$ depicting the in-gap states behavior with respect to time. The dotted/ solid curves closest to zero are associated to  our defined computational states, and the remaining dotted curves are associated to states which are bound to the junction for reasons mentioned in the previous section. Among these four ground states, there is good seperation amongst the energy levels which indicates the possibility of well-behaved gate representation. Here, we include these in-gap, noncomputational states within the calculation of $q(t)$ to quantify whether or not the computational manifold is disturbed by the presence of these lowlying states. This data is plotted on the next row of Figure \ref{fig:TQb_phiprotocol} to demonstrate that they do remain decently situated within the ground state manifold for the duration of the gate operation. In the final subplot, we report the time-dependent diagonal elements of $U(t)$ for the $\phi-$protocol with the remaining off-diagonal elements being approximately zero. For $0<t/T<1/2$, the total effect of $U(t)$ is qualitatively similar to the in-line $\phi-$method system where $U(T/2)\cong \text{exp}(i\pi/2)\mathbb{I}$. However, the subsequent step sends the complex part of the diagonal to zero while sending the real part of each element to $\pm 1$ depending on the state, meaning the total effect is approximately the phase gate,
\begin{gather*}
    U(T)\cong \begin{pmatrix}
        1&0\\
        0&-1
    \end{pmatrix}.
\end{gather*}
The gate calculated in Section \ref{sec:phiShuttle} was simply a global phase, which doesn't do much for real world hardware; however, this protocol does result in a nontrivial gate operation. If $T$ is increased, the states will continue to oscillate through a purely real valued matrix. However, the final condition of the system will only be a global phase factor off from the results reported here.

\vspace{-10pt}
\section{Conclusion: \label{sec:conc}}
\vspace{-10pt}
In summary, we have extended the typical adiabatic computation for non-abelian anyons within rudimentary TSC nanowire systems into a finite-time regime. This work considered the most simplest proposed methods for manipulating MBS which use localized potential and phase differences to shuttle bound states. The core findings for this work are the numerically calculated gate elements presented within Figures \ref{fig:tqbMu} and \ref{fig:TQb_phiprotocol} which showcase the finite-time character of two braiding protocols for a practical platform. In addition to these, we believe our secondary findings captured in Figures \ref{fig:timeDepShuttles_mu} and \ref{fig:timeDepShuttles_phi} provide new insights into the finite-time behavior of the single nanowire cases as well. 

In total, we hope our work can be used to make further judgements regarding topological T-qubits and even more complicated nanowire systems. It is possible to use these results to facilitate the construction of complex meshes of T-qubits and likewise determine their behavior. Additionally, one could combine the methods investigated here into complicated interaction Hamiltonians for discovering gates other than those computed here. These combinations of gate interactions might assist the search for universal quantum gates within topological systems. Work along this vein of ideas will certainly aid the field of TQC by enhancing our collective understanding of the real-world dynamics of nanowires, and we hope our work will provide a stepping stone towards bringing a truely fault tolerant qubit into actuality, making it possible to finally deny the artificial atom its natural inclination to decohere.
\bibliography{main}
\end{document}